\documentclass{emulateapj}

\newcommand{\jms}{J. Mol. Spectrosc.}

\newcommand{\sci}{Science}
\newcommand{\subscript}[1]{\textnormal{\scriptsize{#1}}}
\newcommand{\rstar}{\ensuremath{R_\star}}
\newcommand{\acet}{\ensuremath{\textnormal{C}_2\textnormal{H}_2}}
\newcommand{\diacet}{\ensuremath{\textnormal{C}_4\textnormal{H}_2}}
\newcommand{\odiacet}{\ensuremath{o\textnormal{-C}_4\textnormal{H}_2}}
\newcommand{\triacet}{\ensuremath{\textnormal{C}_6\textnormal{H}_2}}
\newcommand{\tetracet}{\ensuremath{\textnormal{C}_8\textnormal{H}_2}}
\newcommand{\ethylene}{C$_2$H$_4$}
\newcommand{\kms}{km~s$^{-1}$}
\newcommand{\cm}{cm$^{-1}$}
\newcommand{\cmm}{cm$^{-2}$}
\newcommand{\irc}{IRC+10216}
\newcommand{\mlr}{M$_\odot$~yr$^{-1}$}

\begin{document}

\title{Carbon chemistry in \irc: infrared detection of diacetylene}
\shorttitle{Infrared detection of diacetylene in \irc}

\shortauthors{J. P. Fonfr\'{\i}a et al.}
\author{J. P. Fonfr\'ia\altaffilmark{1}}
\author{M. Ag\'undez\altaffilmark{1}}
\author{J. Cernicharo\altaffilmark{1}}
\author{M. J. Richter\altaffilmark{2,4}}
\author{J. H. Lacy\altaffilmark{3,4}}
\affiliation{
  $^1$Molecular Astrophysics Group, Instituto de Ciencia de Materiales de Madrid, CSIC, C/ Sor Juana In\'es de la Cruz, 3, Cantoblanco, 28049, Madrid (Spain); jpablo.fonfria@csic.es\\
  $^2$Physics Dept. - UC Davis, One Shields Ave., Davis, CA 95616 (USA)\\
  $^3$Astronomy Dept., University of Texas, Austin, TX 78712 (USA)
}
\altaffiltext{4}{Visiting Astronomer at the Infrared Telescope Facility, which is operated by the University of Hawaii under contract NNH14CK55B from the National Aeronautics and Space Administration.}

\begin{abstract}
We present the detection of \diacet{} for first time in the envelope of the C-rich AGB star \irc{} based on high spectral resolution mid-IR observations carried out with the Texas Echelon-cross-Echelle Spectrograph (TEXES) mounted on the Infrared Telescope Facility (IRTF).
The obtained spectrum contains 24 narrow absorption features above the detection limit identified as lines of the ro-vibrational \diacet{} band $\nu_6+\nu_8(\sigma_u^+)$.
The analysis of these lines through a ro-vibrational diagram indicates that the column density of \diacet{} is $(2.4\pm 1.5)\times 10^{16}$~cm$^{-2}$.
Diacetylene is distributed in two excitation populations accounting for 20 and 80\% of the total column density and with rotational temperatures of $47\pm 7$ and $420\pm 120$~K, respectively.
This two-folded rotational temperature suggests that the absorbing gas is   located beyond $\simeq 0\farcs4\simeq 20\rstar$ from the star with a noticeable cold contribution outwards from $\simeq 10\arcsec\simeq 500\rstar$.
This outer shell matches up with the place where cyanoacetylenes and carbon chains are known to form due to the action of the Galactic dissociating radiation field on the neutral gas coming from the inner layers of the envelope.
\end{abstract}


\maketitle

\section{Introduction}
\label{sec:introduction}

Evolved stars are known to develop a circumstellar envelope surrounding their central objects.
Around a third part of the total number of molecules discovered in space are present in the envelope of this kind of stars \citep*[e.g.,][]{guelin_1978,guelin_1987,hinkle_1988,bernath_1989,ohishi_1989,bell_1993,cernicharo_1996,cernicharo_2000,cernicharo_2015,anderson_2014,agundez_2014a,agundez_2014b}.
Most of these molecules are formed in the outer shells of the envelopes due to the active chemistry triggered by the radicals and ions that arise after the dissociation of neutral molecules by the external UV radiation field \citep{millar_1994,millar_2000,cernicharo_2004,agundez_2017}.
In particular, the abundances of polyynes (C$_{2n}$H$_2$) and cyanopolyynes (HC$_{2n+1}$N) that have been observed only in the C-rich proto-planetary nebula CRL618 so far can be explained by a photochemical model in which these molecules are formed by chemical reactions involving the radicals C$_{2n}$H and C$_{2n+1}$N, giving raise to a polymerization mechanism that produces carbon-chain molecules \citep{woods_2003,cernicharo_2004}.
The abundance of these molecules decreases as their number of atoms increases depending strongly on the temperature and density of the gas \citep{cernicharo_2004}.

To date, several members of the cyanopolyyne family have been detected in several evolved stars including the very well known Asymptotic Giant Branch star \irc{} \citep*[HC$_{2n+1}$N, $n=0,\ldots,4$;][]{winnewisser_1978,henkel_1985,matthews_1985,guelin_1991}.
Regarding the polyyne family, no member has been detected so far in this source apart from \acet, observed with a column density of $\sim 10^{19}$~\cmm{} \citep{cernicharo_1999,fonfria_2008}.
This non-detection suggests low column densities for other polyynes ($\lesssim 10^{16}$~\cmm), contrarily to what happens in CRL618.
\citet{cernicharo_2001} and \citet{fonfria_2011} clearly detected the \diacet{} and \triacet{} features produced in the photodissociation shells of the envelope of this proto-planetary nebula with column densities $\sim 10^{17}$~\cmm, similar to that of \acet.
This remarkable difference is an effect of the gas density in the photochemical evolution of the envelopes of evolved stars and the photopolimerization of \acet{} and HCN \citep{cernicharo_2004}.

In this paper, we present the first detection of the spectrum of \diacet{} toward the C-rich AGB star \irc.
The observations are described in Section~\ref{sec:observations}.
Section~\ref{sec:results} contains the results of the data analysis.
A discussion about them and their implications in the current chemical scenario of \irc{} can be found in Section~\ref{sec:discussion}.
A brief summary of our work and the final conclusions are in Section~\ref{sec:conclusions}.

\section{Observations}
\label{sec:observations}

\begin{figure*}[!hbt]
  \centering
  \includegraphics[width=\textwidth]{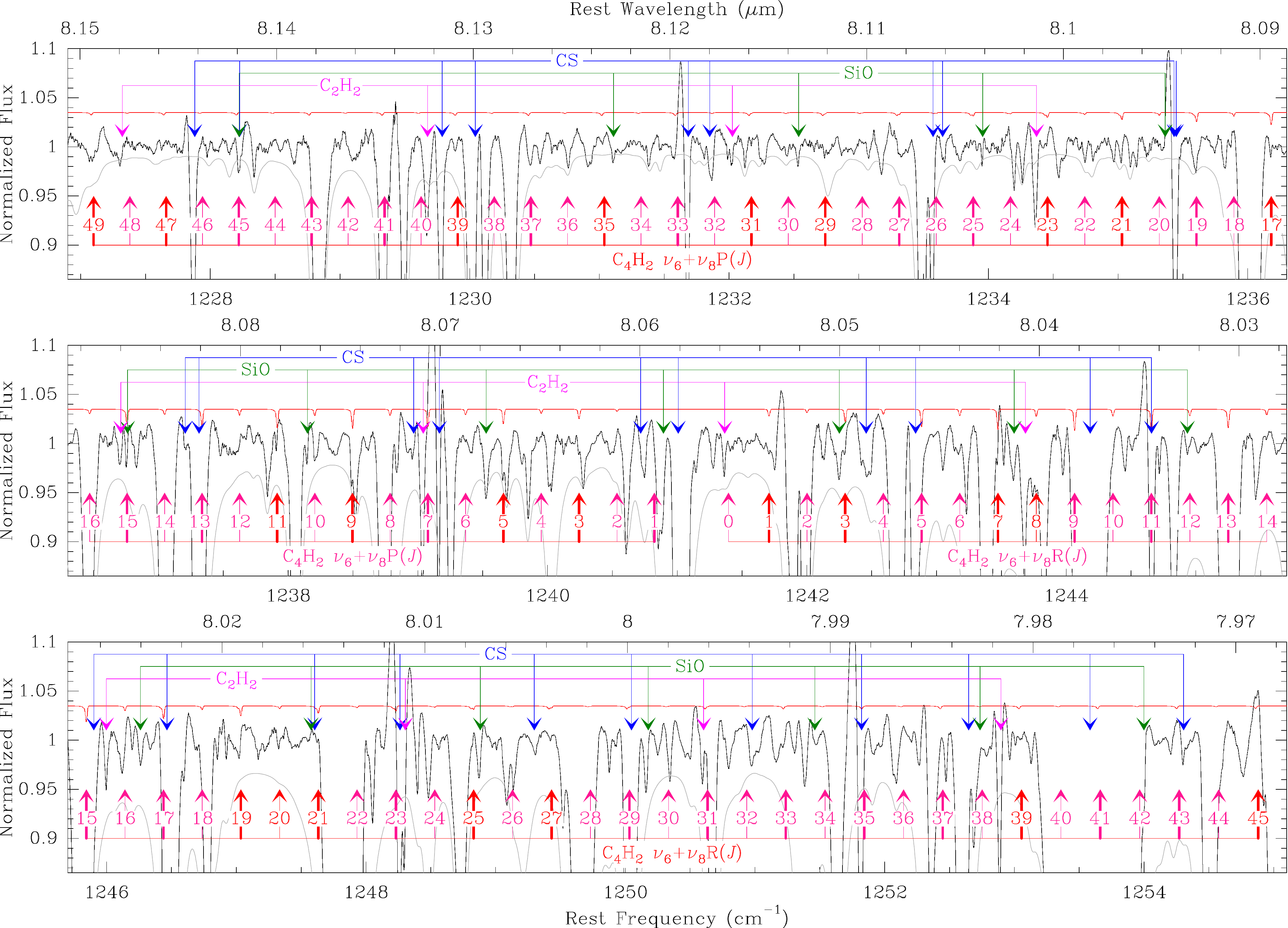}
  \caption{Observed spectrum of \irc{} around $\simeq 1241$~\cm{} against rest frequency (black histogram) containing the $R$ and $P$ branches of band $\nu_6+\nu_8$ of \diacet{} (see text for a definition) and a rough model of the \diacet{} spectrum (red; see Section~\ref{sec:discussion}).
    The telluric transmission is plotted in gray.
    The thickness of the arrows indicates the statistical weight of the line (thick: $o$-\diacet; thin: $p$-\diacet).
    The red arrows point at the identified \diacet{} lines.
    The pink ones, the \diacet{} lines fully blended with other molecular features.}
  \label{fig:f1}
\end{figure*}

The observations were carried out with the Texas Echelon-cross-Echelle Spectrograph (TEXES) \citep{lacy_2002} mounted on the 3~m Infrared Telescope Facility (IRTF) on May 2008.
TEXES was used in its High\_Medium mode with a resolving power of $R\simeq 85,000$, which provides us with a spectral resolution of $\simeq 3-4$~\kms.
\irc{} was nodded along the slit to allow for a better sky subtraction and an efficient on-source integration.
The observations were corrected from the atmosphere with a black body-sky difference.
The data were reduced with the standard TEXES pipeline.
The baseline of each order was removed with an up to 10$^\subscript{th}$ order polynomial fit, taking care of excluding all the features in the spectrum in this process.
The total spectrum was composed of 150 different segments that cover almost completely the spectral range $7.9-9.1~\mu$m.
The part of the spectrum where the \diacet{} lines are found roughly ranges from 8.0 to 8.1~$\mu$m (Figure~\ref{fig:f1}).
The noise RMS is estimated to be $\simeq 0.2\%$ of the continuum.

The spectrum was not corrected from telluric contamination since \irc{} is much brighter at $8~\mu$m than any available calibrator.
The telluric feature identification has been performed by comparing the observations with the Atmospheric TRANsmission (ATRAN)\footnote{\url{https://atran.sofia.usra.edu/cgi-bin/atran/atran.cgi}} model \citep{lord_1992}.
The identification of the features coming from \irc{} has been performed with the aid of the data in the last version of the HIgh-resolution TRANsmission molecular absorption Database (HITRAN)\footnote{\url{http://www.hitran.org}} \citep{rothman_2013}.
The Doppler shift affecting these lines was accurately removed by fitting the strong CS and SiO lines in the observed data.
The analysis of the spectra of these molecules will be published elsewhere.

\section{Results}
\label{sec:results}

\diacet{} is a molecule with 5 stretching modes and 4 doubly-degenerated bending modes.
Only the fundamental modes $\nu_4(\sigma_u^+)$, $\nu_5(\sigma_u^+)$ (stretching bands at $\simeq 3.0$ and $5.0~\mu$m), and the symmetric C$\equiv$C$-$H and C$-$C$\equiv$C bending modes $\nu_8(\pi_u)$ and $\nu_9(\pi_u)$ (bending bands at $\simeq 15.9$ and $45~\mu$m) are infrared active, i.e., the rotational levels in the corresponding excited vibrational states are radiatively connected in the infrared with those of the vibrational ground state by electric dipole transitions.
The anti-symmetric C$\equiv$C$-$H and C$-$C$\equiv$C bending modes $\nu_6(\pi_g)$ and $\nu_7(\pi_g)$ are infrared inactive.
Vibrational modes $\nu_6$ and $\nu_8$ play in \diacet{} the same role than $\nu_4$ and $\nu_5$ in \acet.
The infrared spectrum of \diacet{} comprises additional combination bands such as $\nu_6+\nu_8(\sigma_u^+)$.
The strongest of all these bands are $\nu_4$ and $\nu_8$ followed by $\nu_6+\nu_8$ \citep{khlifi_1995}.
$\nu_8$ is blocked by the strong telluric ozone band around $15~\mu$m and is unavailable from the ground \citep*[but detected from space by][]{cernicharo_2001}.
$\nu_4$ is highly overlapped with strong bands of \acet{} and HCN, very abundant in \irc{} \citep{fonfria_2008}, and other hydrocarbons such as C$_2$H$_6$ also present in the atmosphere.
Thus, looking for lines of band $\nu_6+\nu_8$ from the ground is a very good choice to detect \diacet{} in spite of the absence of the prominent Q branch\footnote{The $R$, $Q$, and $P$ branches of a band are the set of ro-vibrational lines that fulfill that $J_\subscript{up}=J_\subscript{low}+1$, $J_\subscript{up}=J_\subscript{low}$, and $J_\subscript{up}=J_\subscript{low}-1$, respectively.} typical of $\pi-\sigma$ bands that is present, e.g., in the $\nu_8$ band \citep*[e.g.,][]{fonfria_2008}.

We have identified 24 lines of the $R$ and $P$ branches of the \diacet{} band $\nu_6+\nu_8$ above the detection limit.
Most of these lines belong to \odiacet{} due to the spin statistics.
The detected lines are weak and show narrow features with a peak width dominated by the spectral resolution.
They do not show noticeable emission component probably because of the spectrum noise, overlaps with other spectral features, and the existence of several de-excitation routes from the upper vibrational state such as the hot band $\nu_6+\nu_8-\nu_6$, besides other radiation transfer and chemical reasons (see below and Section~\ref{sec:discussion}).
The intensity of their absorption component suggests a column density $\lesssim 10^{16}$~\cmm.
These line profiles are compatible with a molecule formed either in the outer envelope or as close to the star as $\simeq 10-15\rstar$, as it occurs with C$_2$H$_4$ \citep{fonfria_2017}.
C$_2$H$_4$ shows absorption features produced by two different excitation populations compatible with a molecular species arising in the dust formation zone \citep*[$r\lesssim 20\rstar$;][]{fonfria_2008}, where the gas is still being accelerated and the kinetic temperature is above $\simeq 400$~K, and a significant absorption produced in the colder shells of the outer envelope.
This abundance profile implies red-shifted high excitation lines with respect to the low excitation ones.
The velocity of the line peak absorptions of the strongest \diacet{} lines is $-13.3\pm 0.8$~\kms{} with respect to the systemic velocity \citep*[$\simeq -26.5$~\kms;][]{cernicharo_2000}, typical of ro-vibrational lines formed once the terminal gas expansion velocity in \irc{} has been reached \citep*[$\simeq 14.5$~\kms; e.g.,][]{cernicharo_2000,fonfria_2008}.
However, the velocity of the absorption peaks of the weakest line profiles is closer to the systemic velocity, as for \ethylene{} (Section~\ref{sec:discussion}).

\begin{figure}[!hbt]
  \centering
 \includegraphics[width=0.475\textwidth]{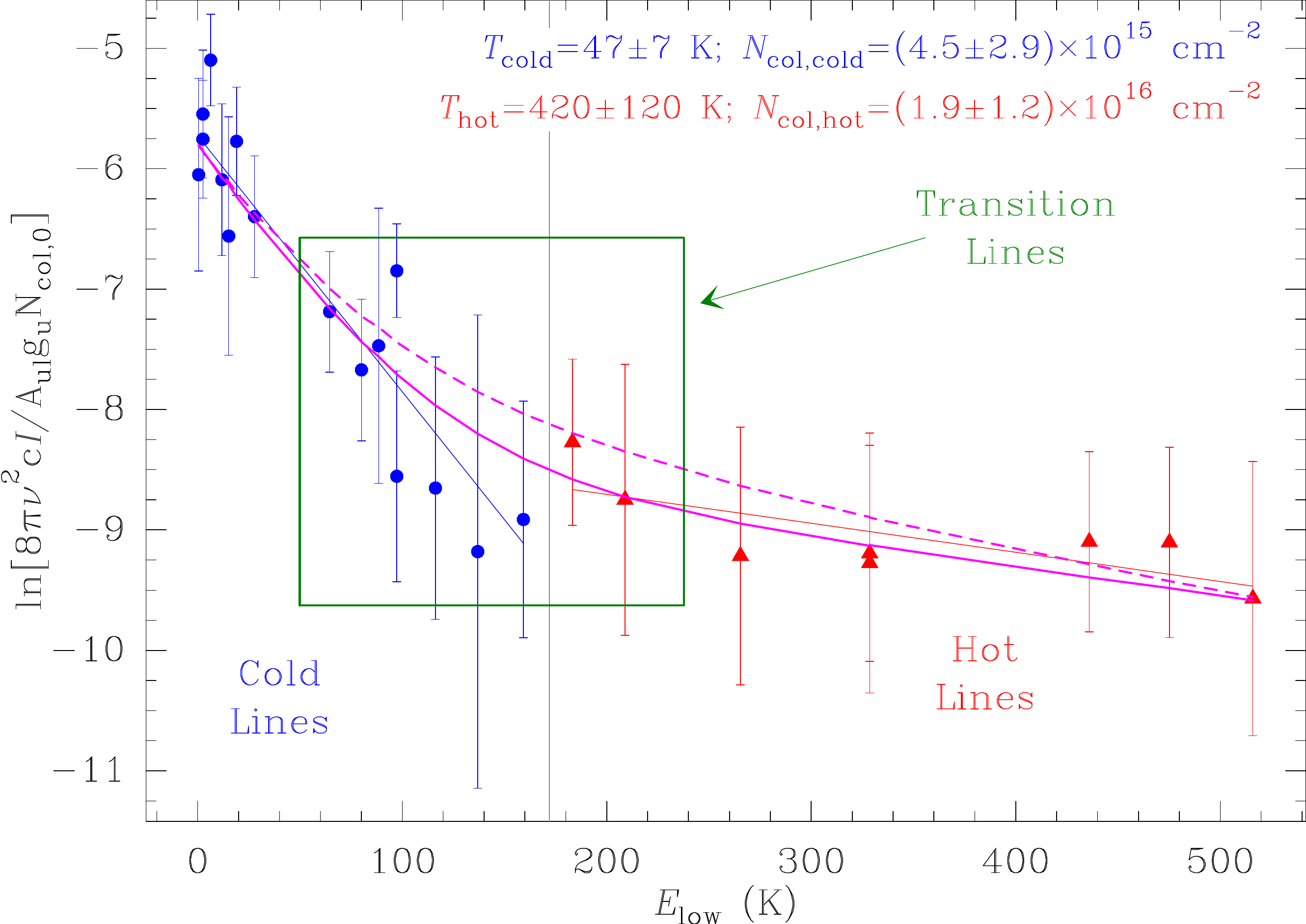}
 \caption{Ro-vibrational diagram of \diacet.
   The data set can be divided into two different groups with different rotational temperatures (red triangles and blue dots for hot and cold lines, respectively).
   The green box contains the lines between both populations (transition lines).
   The blue and red solid straight lines are the fits to the corresponding data sets calculated assuming a weight equal to $1/\sigma^2$, where $\sigma$ is the data uncertainty.
   The magenta curves are derived from two synthetic spectra assuming that \diacet{} is distributed in two isolated shells (solid) and in one thick shell (dashed; see Section~\ref{sec:discussion}).}
  \label{fig:f2}
\end{figure}

The use of a ro-vibrational diagram allows us to estimate the rotational temperature in the vibrational ground state (Fig.~\ref{fig:f2}).
By assuming the same rotational temperature for the excited vibrational states and a given vibrational temperature (see below), we can calculate the total partition function and then obtain the total column density of \diacet.
The \diacet{} abundance is expected to be low so we can consider the observed lines as optically thin.
Thus, the following formula holds for each line:
\begin{equation}
\label{eq:boltzmann}
\ln{\left[\frac{8\pi\nu^2 cI}{A_{ul}g_uN_\subscript{col,0}}\right]\simeq \ln{\left[\frac{N_\subscript{col}}{N_\subscript{col,0}Z}\frac{\theta_s^2}{\theta_s^2+\theta_b^2}\right]}-\frac{hcE_\subscript{low}}{k_\subscript{B}T_\subscript{rot}}}
\end{equation}
where $\nu$ is the rest frequency (\cm), $I$ the integrated absorption, $A_{ul}$ the A-Einstein coefficient (s$^{-1}$), $g_u$ the degeneracy of the upper level, $Z$ the total partition function, $N_\subscript{col}$ the column density (cm$^{-2}$), $E_\subscript{low}$ the energy of the lower ro-vibrational level (\cm), and $T_\subscript{rot}$ the rotational temperature (K).
The integrated absorption was estimated by means of a Gaussian fit chosing the proper baseline.
This baseline was the continuum emission for isolated \diacet{} lines and the profile of a molecular feature around the \diacet{} line if it took part of a blending.
The total partition function was calculated by direct summation over all the available ro-vibrational levels.
The lack of hot bands and of an emission component in the detected lines prevents us to derive the vibrational temperature of \diacet, necessary to calculate the partition function and, thus, the column density.
We have then assumed the vibrational temperature for \diacet{} equals that of the \acet{} band $\nu_4+\nu_5(\sigma_u^+)$ \citep{fonfria_2008}.
$N_\subscript{col,0}=10^{15}$~cm$^{-2}$ is a fixed column density included for convenience to get adimensional arguments for the logarithms.
The factor $\theta_s^2/(\theta_s^2+\theta_b^2)$ is the Point Spread Function filling factor, where $\theta_s$ is the angular size of the \diacet{} absorption and $\theta_b$ results from the quadratic addition of the telescope half power beam width and the atmospheric seeing.
It was $\simeq 0\farcs9$ at $8\mu$m during our observing run.
The size of the \diacet{} absorption can be roughly estimated with our radiative transfer code \citep{fonfria_2008,fonfria_2014} assuming the rotational temperature derived from the ro-vibrational diagram.
Thus, $\theta_s=0\farcs63\pm 0\farcs19$ and $\theta_s^2/(\theta_s^2+\theta_b^2)=0.33\pm 0.13$.
The uncertainties have been calculated propagating the noise RMS of the observed spectrum.

From this diagram, we derive the existence of two different \diacet{} populations (cold and hot) with rotational temperatures of $47\pm 7$ and $420\pm 120$~K, and column densities of $(4.5\pm 2.9)\times 10^{15}$ and $(1.9\pm 1.2)\times 10^{16}$~\cmm, respectively.
The total diacetylene column density is $(2.4\pm 1.5)\times 10^{16}$~\cmm.
This low column density distributed into rotational levels with energies spanning along several hundred of K explains the weakness of the observed lines in the mid-IR.

The lack of a permanent dipole moment of \diacet{} prevents it to have a rotational spectrum.
Therefore, it is rotationally under LTE or very close to it in high density environments but the population of the rotational levels can be significantly affected by the infrared continuum in low density ones.
This effect is difficult to quantify so, to a first approximation, we can place the shell where the observed lines are formed by using the kinetic temperature radial profile recently derived by \citet{guelin_2017} from high spatial resolution data of $^{12}$CO ($T_\subscript{k}\simeq 257(r/0.8)^{-0.675}$~K, if $r\lesssim 15\arcsec$ and $T_\subscript{k}\simeq 35$~K beyond).
Hence, the cold population of \diacet{} is at $10.0^{+2.7}_{-1.8}$~arcsec from the star \citep*[$500^{+140}_{-90}\rstar=\left(1.8^{+0.5}_{-0.3}\right)\times 10^{16}$~cm, if $\rstar=0\farcs02=3.7\times 10^{13}$~cm;][]{ridgway_1988,fonfria_2017} while the hot population arises at $0.40^{+0.30}_{-0.14}$~arcsec from the star ($20^{+15}_{-7}\rstar=\left(7^{+6}_{-3}\right)\times 10^{14}$~cm).
We estimate the \diacet{} abundance with respect to H$_2$ to be $6\times 10^{-7}$ and $8\times 10^{-6}$ for the hot and cold populations, respectively, with an uncertainty roughly of a factor of 2.

\subsection{Searching for other polyynes}

Polyynes \diacet{} and \triacet{} were already found in the proto-planetary nebulae CRL618 in low and high spectral resolution spectra by \citet{cernicharo_2001} and \citet{fonfria_2011}.
However, \tetracet{} is still being searched in space.

Combination bands of molecules such as \triacet{} or \tetracet{} are also centered around $8~\mu$m \citep*[$\nu_8+\nu_{11}$ at 1232.9043~\cm{} and $\nu_{10}+\nu_{14}$ at $\simeq 1229.6$~\cm, respectively;][]{mcnaughton_1991,shindo_2001}.
However, we unsuccessfully looked for features of these molecules in our observations.
Too weak lines with respect to the RMS noise and a covered spectral range too crowded with molecular features are the most probable reasons for this failure.

\section{Discussion}
\label{sec:discussion}

\begin{figure}[!hbt]
  \centering
 \includegraphics[width=0.475\textwidth]{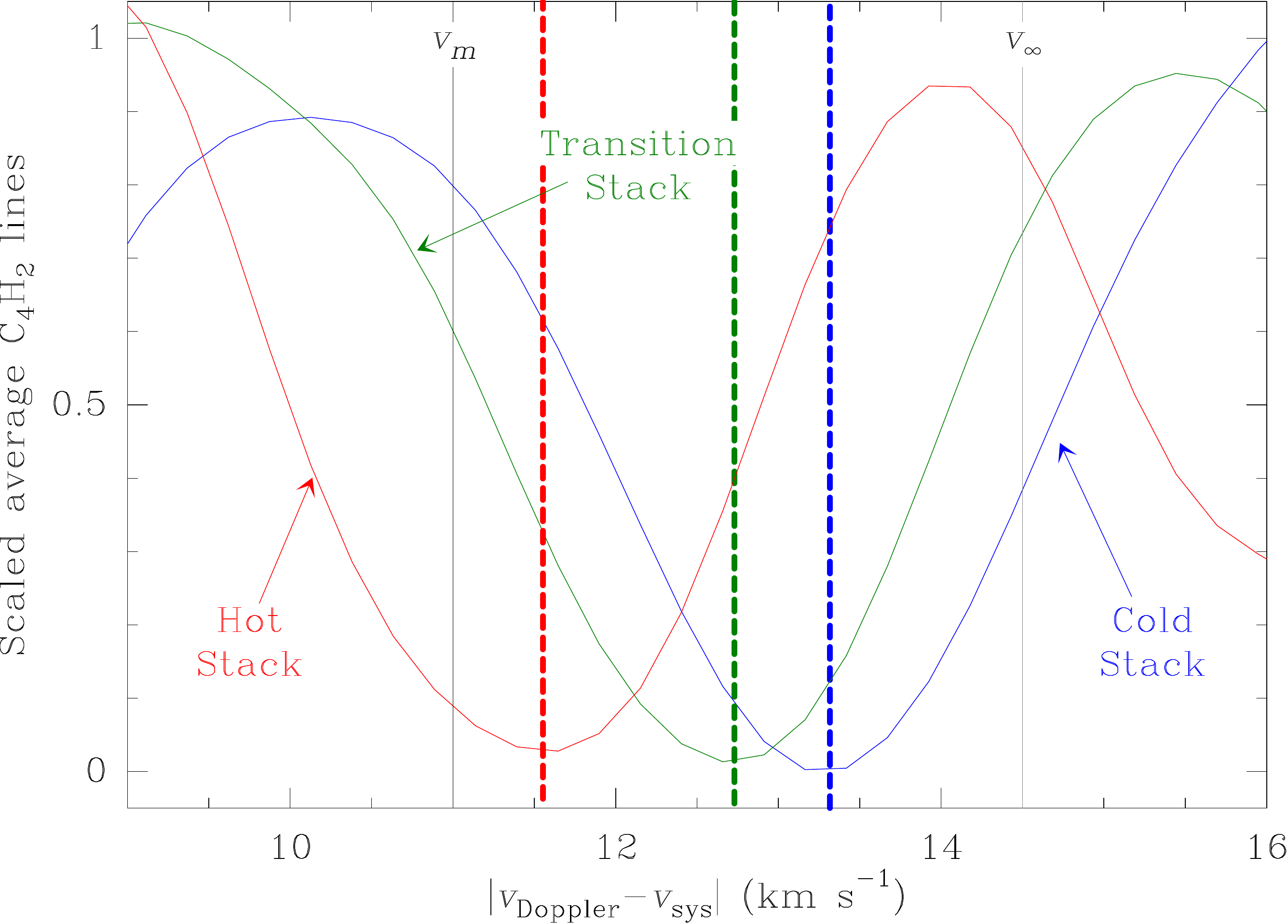}
 \caption{Stacking of lines with different rotational temperatures regarding the gas expansion velocity.
   The lines comprising each stack have been previously scaled.
   The cold and hot stacks involve lines with $E_\subscript{low}\lesssim 40$ and $\gtrsim 240$~K (blue and red).
   The transition stack, plotted in green, comprises lines with rotational temperatures from 47 to 420~K (Fig.~\ref{fig:f2}).
   The velocities $v_m$ and $v_\infty$ are the expansion velocity from 5 to 20\rstar, and the terminal velocity reached beyond 20\rstar{} \citep*[11 and 14.5~\kms; e.g.,][]{fonfria_2015}.}
  \label{fig:f3}
\end{figure}

The preliminary results derived from the ro-vibrational diagram (Section~\ref{sec:results}) suggest that the \diacet{} hot population with a rotational temperature of $\simeq 400$~K arises from regions located around 20\rstar{} from the star, similarly to the case of \ethylene{} \citep{fonfria_2017}.
The gas in this region of the envelope could expand at a lower velocity than the terminal velocity ($\simeq 11$ vs. $\simeq 14.5$~\kms), although it is still under debate \citep*[e.g.,][]{decin_2015,fonfria_2015}.
Adopting this gas expansion velocity field in the dust formation zone, the lines formed at $r\lesssim 20\rstar$ are expected to show a velocity shift of a few \kms{} compared to those formed in the outer shells of the envelope.
To explore this effect, we have divided the observed lines into three different groups, cold ($E_\subscript{low}\lesssim 40$~K), hot ($E_\subscript{low}\gtrsim 240$~K) and transition ($40~\textnormal{K}\lesssim E_\subscript{low}\lesssim 240$~K), stacking the previously scaled lines of each group to improve the S/N ratio and reduce the random shift of the absorption peak due to the spectral noise and the wavelength calibration uncertainties (Fig.~\ref{fig:f3}).
The peak absorption of the cold stack, comprising lines formed in the outer envelope, is shifted $\simeq 1$~\kms{} with respect to the hot stack, formed by averaging the lines supposed to arise in the dust formation zone.
The absorption peak of the transition stack, composed of the lines mostly formed around the acceleration shell at 20\rstar, is placed between both.
This scenario supports the chosen gas expansion velocity profile and the formation of \diacet{} in the dust formation zone, something unpredicted by the most commonly accepted photochemical models \citep{millar_1994,millar_2000,agundez_2017}.

The observed \diacet{} lines can be roughly modeled with the code employed by \citet{fonfria_2017} to model the \ethylene{} spectrum toward \irc.
In this model we adopted (1) the same kinetic and vibrational temperatures used during the analysis of the observational ro-vibrational diagram (Fig.~\ref{fig:f2}), (2) a mass-loss rate of $2.7\times 10^{-5}$~\mlr{} \citep{guelin_2017}, (3) a distance of 123~pc \citep{groenewegen_2012}, and (4) the gas expansion velocity proposed by \citet{fonfria_2015}, i.e., $1+2.5(r/\rstar-1)$~\kms{} if $1\le r/\rstar<5$, 11~\kms{} if $5\le r/\rstar<20$, and 14.5~\kms{} if $r/\rstar\ge 20$.
We considered two different models compatible with the diacetylene abundance distribution derived in Section~\ref{sec:results} to explore how its variation affects the ro-vibrational diagram: (1) \diacet{} is distributed in two isolated shells ranging from 15 to 20\rstar{} and from 400 to 1000\rstar, and (2) \diacet{} is formed at 15\rstar{} with a constant abundance up to 400\rstar, adopting another constant value beyond.
We chose a distance of 15\rstar{} as the inner radius of the abundance distribution because a value of 20\rstar{} results in ro-vibrational diagrams with a significantly steeper slope at high $E_\subscript{low}$ than the observed data suggest.
Noticeable emission components arose in the synthetic lines after assuming shorter distances, something unobserved in our spectrum.
The results are plotted in Fig.~\ref{fig:f2} as the solid and dashed magenta curves.
Both models are compatible with the observational ro-vibrational diagram but the data seems to be better reproduced with model 1 (plotted in Fig.~\ref{fig:f1}).
However, the transition lines ($E_\subscript{low}\simeq 100-200$~K) can also be slightly influenced by a low abundance \diacet{} contribution with a rotational temperature of a few hundred K located between $\simeq 20$ and 400\rstar.

\begin{figure}[!hbt]
  \centering  \includegraphics[width=0.475\textwidth]{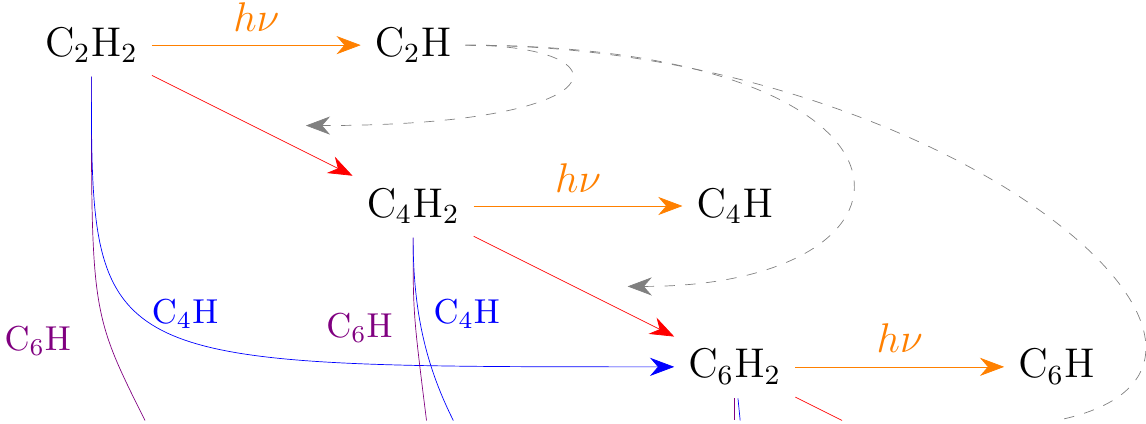}
  \caption{Diagram showing the reaction network involved in the polymerization of polyynes from the dissociation of \acet{} \citep{agundez_2017}.
The solid arrows indicate reactions involving the molecules at the beginning of the arrows and by the arrows (with the same color) that give the products at their ends (e.g., C$_2$H$_2$+C$_4$H$\to$C$_6$H$_2$+H in blue).
The dashed arrows mean that the molecules at their beginnings are involved in the processes to which the arrows point at (e.g., C$_2$H$_2$+C$_2$H$\to$C$_4$H$_2$+H).
All the reactions in this diagram give a free H atom as a product.}
  \label{fig:f4}
\end{figure}

\begin{figure}[!hbt]
  \centering
  \includegraphics[width=0.475\textwidth]{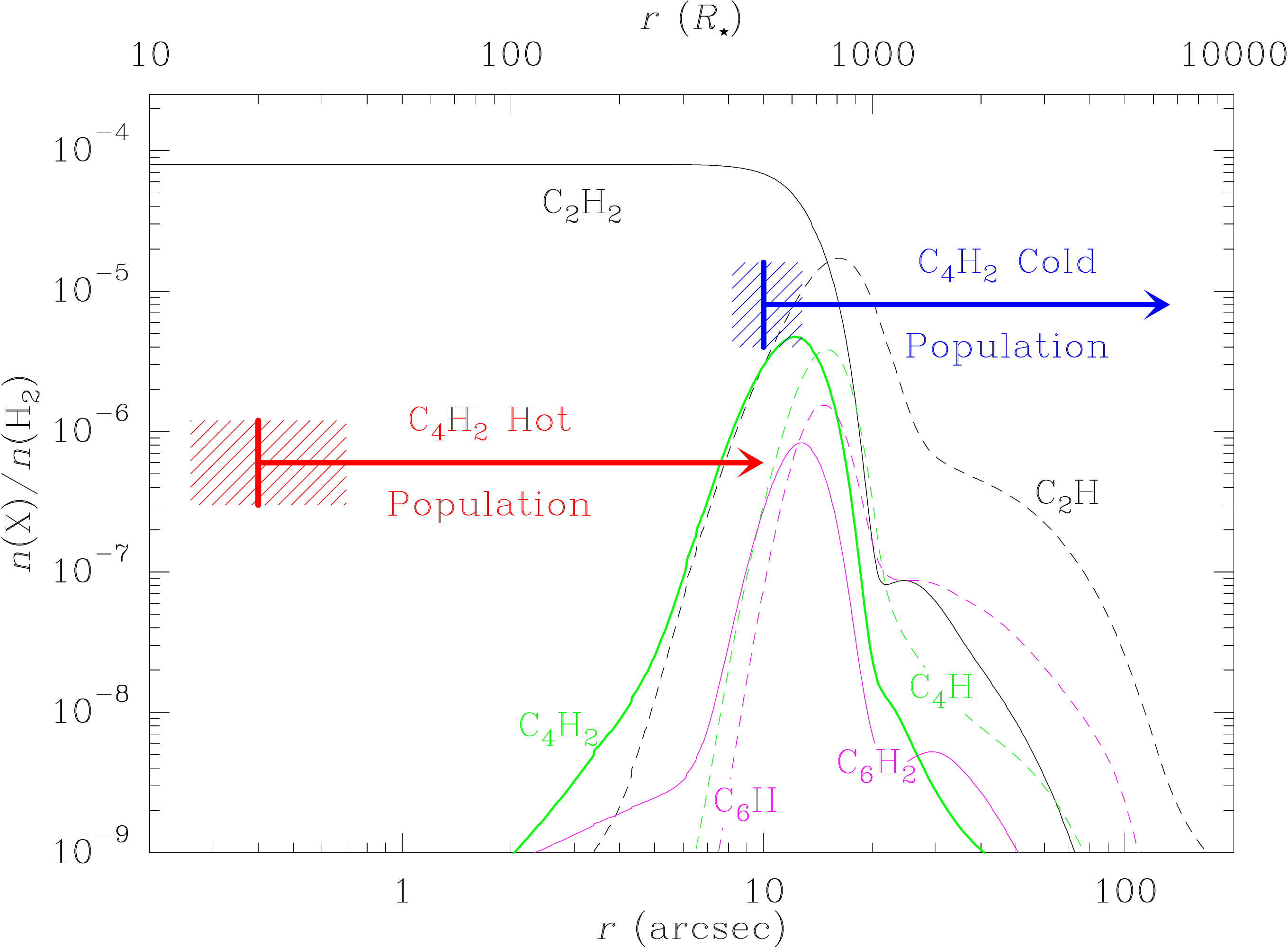}
\caption{Abundances with respect to H$_2$ of polyynes (C$_{2n}$H$_2$) and carbon chains (C$_{2n}$H) in the outer envelope of IRC+10216.
The abundance of \diacet{} derived in the current work is included as two horizontal arrows for the detected populations (hot in red and cold in blue) to highlight that we know where the shells with different population start but we ignore where they end.
The coordinate in the vertical axis of these arrows is the \diacet{} abundance in each shell.
The vertical lines at the beginning of each arrow indicate the positions of the inner boundaries of the shells with the inferred \diacet{} populations.
The solid and dashed curves are molecular abundances calculated by \citet{agundez_2017}.
The hatched regions indicate the uncertainties of the position of the inner boundaries of the \diacet{} shells and of its abundance in them.}
  \label{fig:f5}
\end{figure}

The photochemical models for the outer envelope reproduce reasonably well the abundances of the molecules formed in the shells of C-rich evolved stars such as \irc{} irradiated by dissociating radiation, in particular of cyanopolyynes HC$_{2n+1}$N \citep{millar_1994,millar_2000,cernicharo_2004,agundez_2010,agundez_2017}.
These models predict that \diacet{} (and larger polyynes) arises mostly due to the reaction \acet+C$_2$H$\to$\diacet+H that can happen after the formation of C$_2$H, which results from the photodissociation of \acet{} (Fig.~\ref{fig:f4}).
This polymerization process produces polyynes of increasing length, as can be seen in Fig.~\ref{fig:f5}, where we show the radial abundance profiles calculated with the photochemical model by \citet{agundez_2017}.
Briefly, this model calculates the chemical evolution of the isotropically expanding gas around a cold star starting at $\simeq 5\rstar$ from its center, where the considered parent molecules (H$_2$, CO, C$_2$H$_2$, CH$_4$, C$_2$H$_4$, H$_2$O, N$_2$, HCN, NH$_3$, CS, H$_2$S, SiS, SiO, SiH$_4$, PH$_3$, and HCP) were supposed to be already formed.
To reproduce their ALMA observations of carbon chains in \irc, the authors assumed a smooth envelope externally illuminated by the local UV radiation field of \citet{draine_1978} and a ratio $N_\subscript{H}/A_V$ 1.5 times lower than the canonical value of $1.87\times 10^{21}$~cm$^{-2}$~mag$^{-1}$, derived by \citet{bohlin_1978} for the local Interstellar Medium.
This model also takes into account the molecular ionization process triggered by cosmic-rays, known to be significant not only for H$_2$ but for C$_2$H$_2$ as well \citep{gredel_1989}.
The authors adopted a large network of around 8,300 chemical reactions taken mainly from the literature on gas-phase chemical kinetics and the UMIST and KIDA databases \citep{mcelroy_2013,wakelam_2015}.
More details about the model can be found in \citet{agundez_2017}.

The predicted column density associated to the cold population ($\simeq 1.7\times 10^{15}$~\cmm) compared to our estimate ($(4.5\pm 2.9)\times 10^{15}$~\cmm) is particularly satisfactory, lying inside the $1\sigma$ error interval.
This means that the predicted column density of C$_4$H, about half the \diacet{} column density, is well estimated and its spectrum could be detected in the infrared.
The model also predicts fairly well the position of the shell where \diacet{} arises and its abundance, which is compatible with our estimate regarding the cold population within a factor 2.
Nevertheless, this does not occur with the abundance of the hot population, that is several orders of magnitude higher than the model prediction in the shell where \diacet{} forms.
It is noteworthy that, contrarily to what occurs with the column density, which is lower for the cold \diacet{} population with respect to the hot one, the abundance with respect to H$_2$ shown in Fig.~\ref{fig:f5} is higher in the outer envelope than in the inner envelope.
This apparent incompatibility is explained by the fact that the gas density in the outer shells ($\simeq 500\rstar$) is well below its value at $\simeq 20\rstar$ from the star.

The disagreement in the abundance of the hot population found between the observations and the model results can be explained in two different ways.
First, the high excitation \diacet{} lines in the spectrum could be blended with stronger unidentified lines.
However, the intensity of these lines is compatible with the lower excitation lines (Fig.~\ref{fig:f2}), something that would not happen if they were from other molecules.
This leads us to the second way, which suggests that the photochemistry model underestimates the abundance of \diacet{} in the inner envelope.
This would be in line with the idea that the dissociating external radiation field can reach shells significantly closer to the star than usually accepted.
It is observationally supported by the works based on data acquired in the visible and the FUV by \citet{leao_2006}, \citet{kim_2015}, and \citet{matthews_2015}.
Since the bulk opacity for this external dissociating radiation field is the dusty component of the envelope, dissociating photons can go deeper into the envelope if dust grains are inhomogeneously distributed in clumpy shells.
Several authors have demonstrated that the circumstellar chemistry can be significantly modified if a higher density, higher temperature photochemistry and clumpiness are considered \citep{woods_2003,redman_2003,cernicharo_2004,agundez_2010}.
In this scenario, the abundance of \diacet{} is naturally explained at the same time than the existence of H$_2$O in the envelope of a C-rich star \citep{melnick_2001,agundez_2006,agundez_2010,neufeld_2013} and the discovery of vibrationally excited C$_4$H and \ethylene{} as close to the star as $\simeq 10\rstar$ \citep{yamamoto_1987,cooksy_2015,fonfria_2017}.

\section{Summary and conclusions}
\label{sec:conclusions}

In this paper, we have presented for first time 24 features of the \diacet{} fundamental band $\nu_6+\nu_8(\sigma_u^+)$ observed with a high spectral resolution ($R\simeq 85,000$) towards the C-rich star \irc{} with the TEXES spectrograph mounted on the 3~m telescope IRTF.
From the analysis of this spectrum, we conclude that:
\begin{itemize}
\item There are two \diacet{} populations with different rotational temperatures ($420\pm 120$ and $47\pm 7$~K).
  We estimate that these rotational temperatures are typical of shells at $\simeq 0\farcs4\simeq 20\rstar\simeq 7\times 10^{14}$~cm and $\simeq 10\arcsec\simeq 500\rstar\simeq 1.8\times 10^{16}$~cm from the central star.
\item The total \diacet{} column density is $(2.4\pm 1.5)\times 10^{16}$~\cmm.
  Only about 20\% of it is located at the outer envelope, where the external dissociating radiation field is able to dissociate parent molecules coming from the inner layers of the envelope.
  The rest (80\%) corresponds to \diacet{} formed in the dust formation zone ($r\lesssim 20\rstar$).
\item The underestimation of the \diacet{} abundance predicted by our photochemical model suggest that the molecules in the envelope are photodissociated in shells closer to the star than is commonly assumed.
  The easiest scenario in which this could happen would involve a clumpy outer envelope where the dust grains density undergoes significant variations between different places.
\end{itemize}

\section*{Acknowledgements}

We thank the anonymous referee for his/her comments about the manuscript.
Development of TEXES was supported by grants from the NSF and USRA.
The research leading to these results has received funding support from the European Research Council under the European Union's Seventh Framework Program (FP/2007-2013) / ERC Grant Agreement n. 610256 NANOCOSMOS.
JC and MA thank Spanish MINECO through grants AYA2012-32032 and AYA2016-75066-C-1-P.
MA also thanks funding support from the Ram\'on y Cajal program of Spanish MINECO (RyC-2014-16277).

\facility{IRTF(TEXES)}

\end{document}